\documentclass{aastex631}
\usepackage{amsmath}
\usepackage{multirow}

\newcommand\nasagoddard{
NASA Goddard Space Flight Center, Greenbelt, Maryland, USA
}

\begin{document}
\title{HPIC: The Habitable Worlds Observatory Preliminary Input Catalog}
\author{Noah W. Tuchow}
\affil{\nasagoddard}

\author{Chris Stark}
\affil{\nasagoddard}

\author{Eric Mamajek}
\affil{Jet Propulsion Laboratory, California Institute of Technology, 4800 Oak Grove Drive, Pasadena, CA 91109, USA}
\affil{Department of Physics and Astronomy, University of Rochester, Rochester, NY 14627-0171, USA}

\begin{abstract}

{The Habitable Worlds Observatory Preliminary Input Catalog (HPIC) is a list}
of $\sim$13,000 nearby bright stars that will be potential targets for the Habitable Worlds Observatory (HWO) in its search for Earth-sized planets around Sun-like stars. {We construct this target list using} the TESS and Gaia DR3 catalogs, and develop an automated pipeline to compile stellar measurements and derived astrophysical properties for all stars. We benchmark the stellar properties in the HPIC relative to those of the manually curated ExEP HWO Precursor Science Stars list and find that, for the 164 best targets for exo-Earth direct imaging, our stellar properties are consistent. We demonstrate the utility of the HPIC by using it as an input for yield calculations to predict the science output of various mission designs including those with larger telescope diameters and those focused on other planet types besides Earth analogs, such as Jupiter-mass planets. The breadth and completeness of the HPIC is essential for accurate HWO mission trade studies, and it will be useful for other exoplanet studies and general astrophysics studying the population of bright nearby stars.

\end{abstract}

\section{Introduction}

In the coming decades, astronomers aim to achieve the ability to directly image Earth-sized planets in the habitable zones of Sun-like stars. With advances in coronagraph and telescope technologies, we aspire to obtain atmospheric spectra of Earth-like planets and infer whether they could be habitable or even exhibit biosignature gases.
Given the potential of future instruments to meet these goals, one of the top recommendations of the Astro2020 Decadal Survey is a mission to directly image Earth-sized planets in the habitable zones of their stars and characterize their spectra \citep{DecadalSurvey, HabEx_Final_report,LUVOIR_final_report}. The Decadal Survey recommends a roughly 6m inscribed diameter space telescope, with wide wavelength coverage from the ultraviolet to the infrared and a next-generation coronagraph able to block a star's light while preserving the light of its planets. This proposed telescope is now being studied as a future NASA flagship mission concept and is referred to as the Habitable Worlds Observatory (HWO).
This mission is driven by the science goals put forth in the Decadal Survey, {one of which is the detection and characterization of} roughly 25 earth sized planets in the habitable zones of sun-like stars. 
Obtaining a sufficient sample size of Earth-like planets will allow us to constrain their occurrence rates and place empirical and statistical constraints on the inner and outer boundaries of the region where planets with surface liquid water can be found, testing the concept of the habitable zone \citep{bean2017,lustigyaeger2022}.

We would like to obtain a sample of {at} least 25 Earth analogs with HWO, but directly imaging any planets at these separations and contrasts is no easy task. Planets are inherently very faint compared to their host stars, and those in the habitable zone have sub-arcsecond angular separations from their stars, making them difficult to resolve separately. 
To directly image these planets, one needs to block out the contaminating light of the host stars while preserving the faint signal of the planets. For Earth-like planets, which are roughly 10 billion times fainter than their host stars in reflected light, this would require an advanced coronagraph and telescope designed specifically to accommodate these science requirements. 
Even with future instruments, the direct imaging of Earth-like planets will be limited to nearby bright stars. For these stars, planets would be bright enough in reflected light to not require excessive exposure times, and they would have large enough angular separations from their stars to fall outside the inner working angle of the coronagraph. 
Given that Earth-like planets are incredibly faint, the number of stars a space-based direct imaging mission would be able to observe will be limited by the exposure time necessary to detect and characterize planets.

At this stage in the precursor studies for HWO, trade studies are required to determine whether proposed mission design concepts will be able to meet HWO's science goals. Yield calculations will play a key role in these trade studies by estimating the number of planets that a given architecture will be able to observe \citep{brown2005,Savransky2010,Stark2015,Stark2019,morgan2019}. Fundamentally, exoplanet yield calculations work by ingesting a multitude of inputs describing the astrophysical universe around us, then simulating the performance of an observatory as it executes an exoplanet survey within that ``universe." 

The most fundamental astrophysical input to yield calculations is a catalog of nearby stars with accurate stellar properties. From this catalog, yield codes can optimally select stars that are well-suited for the mission in order to maximize yield \citep{Stark2014}. Many aspects of the star can impact this optimization. For example, distance and luminosity determine the angular extent of the habitable zone on the sky, stellar radius can affect the coronagraph’s raw contrast, mass sets the orbital period of planets and thus cadence of observations, and companion stars (even outside of the instrument’s field of view) can cause problematic stray light. 
Errors in the stellar input catalog can therefore impact yield calculations. 
\citet{Stark2019} found that the net yield of an exo-Earth survey varied by only $5\%$ depending on one's choice of target list. Therefore, relatively rudimentary stellar catalogs were sufficient for the early yield calculations used in the LUVOIR and HabEx mission concept studies. Going forward, we will desire a more precise input catalog, allowing us to achieve a higher degree of accuracy in expected exposure times and to better identify the population of high priority stars for future study. 
Critically, the catalog must also be sufficiently large and complete. To optimally match stars to the mission being studied, the catalog must give the yield code the flexibility needed to maximize yield by providing more stars than what are ultimately selected. An incomplete input catalog could lead to inaccurate trade studies, or the inability to adequately study certain science cases beyond a survey for potentially Earth-like planets.

Multiple input catalogs have been developed to study exoplanet direct imaging with a future telescope like HWO. While all of these catalogs have been extremely useful in their own right, each has important limitations. ExoCat was very useful for probe studies that would be limited to stars interior to 30 pc, but it was developed nearly a decade ago, was curated by hand, and has not been updated with Gaia distances \citep{turnbull2015}. The AYO input catalog, used in the yield calculations for LUVOIR and HabEx, was developed to extend to 50 pc and used Gaia DR2 distances, but was still based on the Hipparcos catalog, was incomplete starting at $V\sim8$, and is known to have some roughly approximated stellar parameters \citep{Stark2019}. The recent HWO Mission Stars List released by the NASA Exoplanet Exploration Program (ExEP) office provides up-to-date, precise stellar properties for the best targets for exoplanet direct imaging \citep{mamajek2023}. This curated list is immensely useful in providing host star properties for the highest priority HWO stars, but it is limited to $\sim$160 stars, impairing the ability of yield codes to optimally select target stars. \citet{mamajek2023} acknowledge that trade studies focused on varying aspects of the mission architecture such as increasing aperture size or considering different values for $\eta_\oplus$ (the occurence rate of Earth-like planets),  will run into the limitations of the target list, and could benefit from a larger input catalog. 

In this paper, we construct a new target list for use in precursor studies for the Habitable Worlds Observatory. Our list, which we call the Habitable Worlds Observatory Preliminary Input Catalog (HPIC), focuses on gathering and characterizing the sample of nearby bright stars that would be able to host observable planets via space-based direct imaging. Unlike previous target lists, the HPIC is not restricted to only the search for Earth analogs, but can also be used for planning missions to image exoplanets of different types, such as those with larger separations and larger masses. The list of target stars that we are developing is preliminary in nature, and, given that telescope architectures have not been settled on, it is designed to be agnostic of mission design. In later stages of mission development, once a design for HWO is finalized, a definitive input catalog for its direct imaging survey will need to be created. However, regardless of the final design of HWO, we can be confident that the stars surveyed in its direct imaging survey will be included in our preliminary input catalog.
This is because there are only a few thousand stars that are bright and near enough to be good targets for direct imaging. While our knowledge of the stellar properties and binarity of these stars will continue to evolve and new faint stars will be discovered, it is highly unlikely that new bright ($V<6$) stars in the solar neighborhood will be discovered prior to the launch of HWO (estimated for the 2040s). 
We expect the HPIC target list to be useful for other upcoming new great observatories as well, not just HWO. While our list is focused on determining the best targets for space-based direct imaging, any future mission focused on studying the sample of bright stars (less than 12th magnitude) in the solar neighborhood, or searching for planets around them, will benefit from the HPIC. 

Our paper is organised as follows.
In Section \ref{methods_section} we describe the methodology for how our input catalog is constructed. We detail the selection of objects for our list, our methodology for gathering and computing stellar properties, and our tests to ensure the {reliability} of these properties. Then, in Section \ref{yield_section}, we demonstrate the utility of the HPIC by using it as the input for exoplanet yield calculations.

\section{Constructing the HPIC}
\label{methods_section}
\subsection{{Selecting Stars for Direct Imaging}}
\label{considerations_section}

{To construct an input catalog for future space-based direct imaging missions such as HWO, one does not need to start from scratch. Direct imaging of Earth-sized planets is feasible only for the population of bright nearby stars, and such stars have been surveyed by several past and ongoing missions. In particular, the target lists for the TESS and Gaia missions are useful starting points when constructing an input catalog for HWO.
The TESS Input Catalog (TIC) contains the observed and derived properties as well as cross-matching information for the list of 1.7 billion objects used to plan the TESS mission's survey for transiting exoplanets \citep{TESS_overview,TIC_DOI, stassun2018,stassun2019,paegert2021}. 
Gaia Data Release 3 contains astrometric data for 1.8 billion objects, and includes precise measurements of parallaxes, distances, proper motions and photometry, as well as astrophysical properties derived by multiple automated pipelines \citep{prusti2016,gaiadr3}.}

\begin{figure}
    \centering
    \plotone{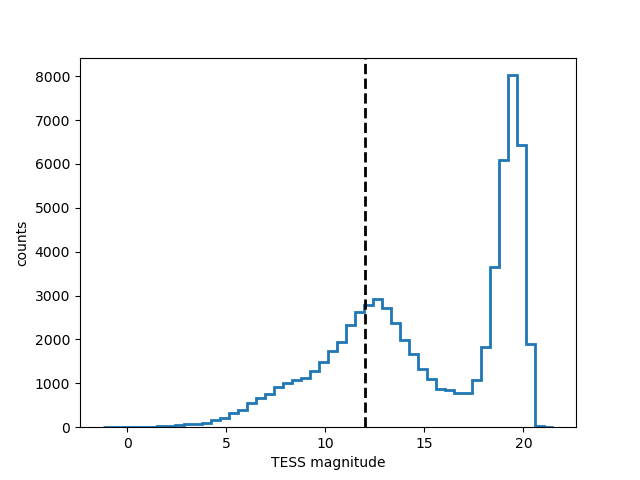}
    \caption{Distribution of TESS magnitudes for TIC objects within 50 pc. The black vertical line shows our magnitude cutoff at 12 which maintains volume completeness.}
    \label{T_hist}
\end{figure}

{
Together, with over a billion stars, the TESS and Gaia target lists
are much too large for a direct imaging survey. To reduce the catalog size and omit stars that are not feasible targets for any direct imaging mission within the trade space, we make two reasonable cuts to the catalog.}
{First, we implement a distance cutoff of 50 pc, beyond which the angular size of the habitable zone will fall within the coronagraph's inner working angle.  At 50 pc, a planet at 1 AU would have an angular separation of just 20 milliarcseconds.  Even for an idealized coronagraph with a $1 \lambda/D$ inner working angle, an 8m diameter telescope operating at 1 micron would be limited to separations greater than 26 mas.}
{Second, we apply an apparent magnitude cutoff to exclude sources that are too faint and would require unrealistically long exposure times to detect planets in reflected light. Though trimming the catalog at 8th magnitude would likely contain all targets for which exo-Earth direct imaging would be feasible, the HPIC is designed to extend to fainter magnitudes to allow for surveys of other planet types and to account for mission architectures spanning the range of the trade space. }
{
We adopt a 12th magnitude cutoff, the faintest our catalog can extend to while still ensuring that the catalog is volume complete. As shown in the histogram of magnitudes of TIC objects in Figure \ref{T_hist}, the number of objects begins to decrease at $T$ magnitudes of $\sim$12.5, indicating that beyond that magnitude the TIC is no longer volume complete.}

\subsection{{Catalog Construction {Pipeline}}}

\begin{figure}
    \centering
    \plotone{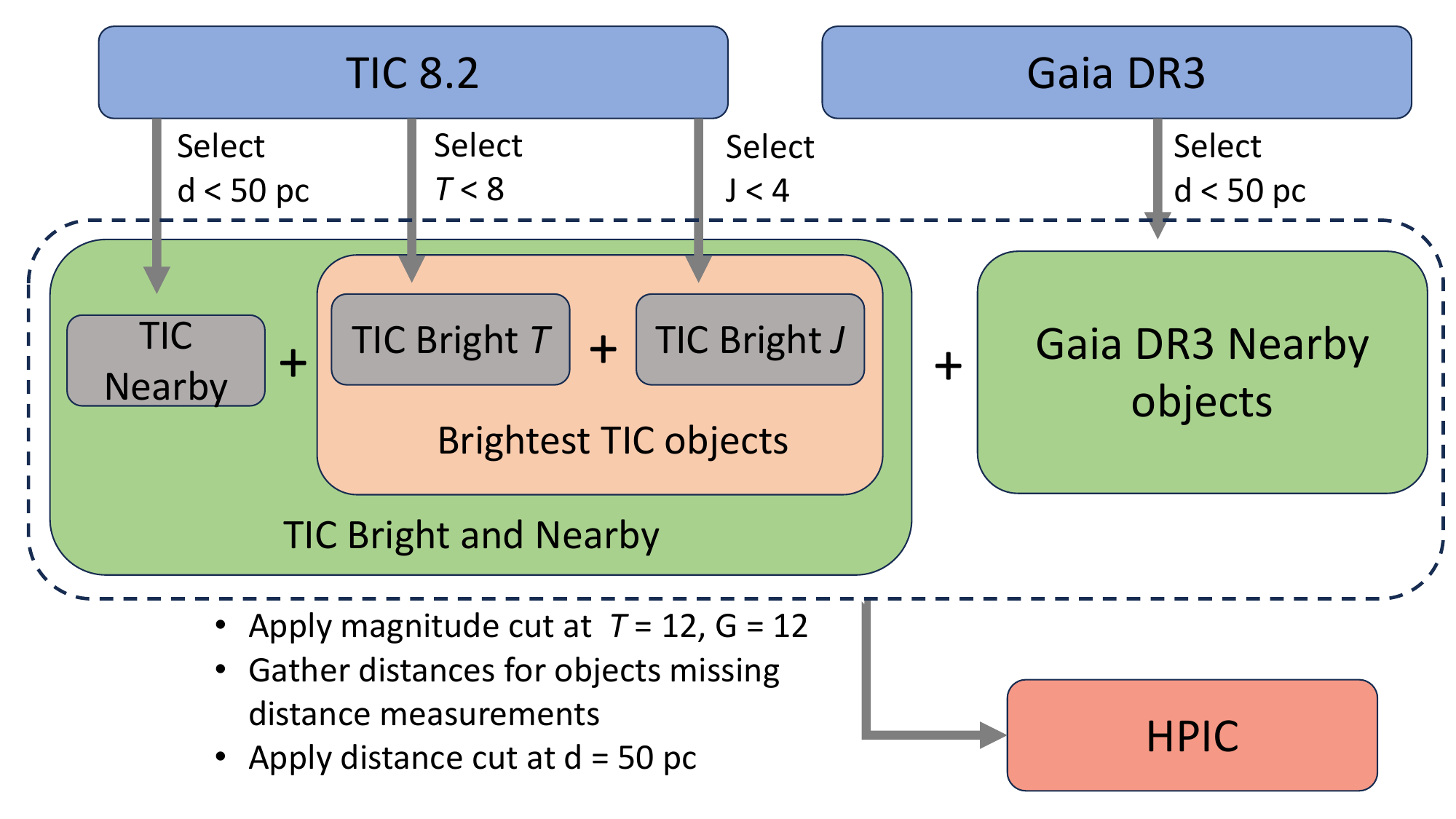}
    \caption{{Diagram illustrating the procedure used to construct the HPIC.}}
    \label{HPIC_diagram}
\end{figure}

Whereas previous catalogs of direct imaging target stars were largely developed via manual curation, {for the HPIC} we have created a pipeline to automate our target list generation. This will allow for simple updates as additional data sets are released in the future, and there will be no ambiguity on how stars were selected and how their stellar properties were calculated. Here we detail the procedure for constructing our target list and the pipeline we have developed {(illustrated in Figure \ref{HPIC_diagram})}.

{To form the foundation of the HPIC, we start with the target lists of the TESS and Gaia missions as described in the previous section.
Our first step is to query the TIC 8.2, hosted on the CDS VizieR service \citep{vizier,TIC_DOI}, for all objects with distances less than 50 pc. 
Selecting all nearby TIC objects with distances less than 50 pc yields the majority of stars that are good candidates for direct imaging, but there are a few hundred objects in the TIC without distances which may still be good targets, such as binaries without accurate parallaxes. To make sure we include these objects in our list, we also include all bright TIC objects, with or without distance measurements. While ideally we could obtain the brightest stars just by selecting based on TESS magnitude, we are faced with the problem that a few dozen of the brightest TIC objects appear to have incorrect estimates of their $T$ magnitudes. This is due to the fact that $T$ magnitudes are often calculated using Gaia photometry, but the brightest objects were too bright to be observed by the Gaia mission and their $T$ magnitudes may be erroneously computed using a faint background Gaia source. In these cases, we select objects based on their 2MASS $J$ magnitudes instead of incorrect $T$ magnitudes. To obtain the brightest TIC objects we make two separate queries: we select objects with $T<8$ to obtain the vast majority of bright stars, and we select objects with $J<4$ to ensure the very brightest objects without accurate $T$ magnitudes are included in our catalog.
}

{We now have two subsets of the TIC: all the objects with distances less than 50 pc and all the brightest TIC objects ($T < 8$ or $J < 4$).
We then take the union of these two subsets of the TIC, removing objects with duplicate TIC identifiers. We remove objects in the resulting combined list with distances in the TIC greater than 50 pc, but keep objects which do not have distances provided. In Section \ref{distance_section} we discuss how we obtain distance estimates for these objects and how objects are removed from our catalog if their newly obtained distance exceeds 50 pc.}

{Apart from the bright and nearby objects in the TIC, there are several hundred nearby objects observed by Gaia DR3 that are not included in the TIC. Most of these are fainter field stars and newly resolved binaries.
To identify these objects we first query the \citet{bailerjones2021} catalog of Bayesian distance estimates for Gaia DR3 sources, selecting all objects with distances less than $50$ pc. Within this list of nearby Gaia objects there are objects that are not in the TIC, and to identify them we need to remove objects that are redundant with TIC sources. }

{To find which objects are the same and remove duplicates, we identify the corresponding Gaia DR3 identifiers for all objects in our subset of the TIC. 
The TESS input catalog was constructed using Gaia DR2, but has not yet been updated for Gaia DR3 \citep{stassun2019,paegert2021}. As such, it includes the Gaia DR2 identifiers for most objects, but does not contain Gaia DR3 identifiers\footnote{Identifiers of Gaia objects in data releases 2 and 3 are not necessarily the same. In many cases, a given Gaia DR2 identifier can correspond to multiple DR3 objects due to newly resolved binary companions or faint background objects that previously weren't resolved separately.}. Fortunately the Gaia team has done extensive cross-matching between DR2 and DR3, and includes these results in the dr2$\_$neighbourhood table on the Gaia Archive \citep{gaiadr3,torra2021}. Using this table for every object in the TIC with a Gaia DR2 identifier, we select the DR3 object that is the closest match in terms of both position and $G$ magnitude. For TIC objects without Gaia DR2 identifiers, we  perform a positional cross-match with Gaia DR3 using the CDS X-Match service \citep{boch2012}. For the list of Gaia DR3 sources within 10 arcseconds of an object, we select the one that is closest in terms of angular separation and TESS magnitude.  Finally, for Gaia DR3 objects which lack a corresponding object in the TIC, we use the same DR2-DR3 neighborhood table as we have discussed earlier to find DR2 identifiers corresponding to each object. This is necessary as many external catalogs we use in later sections identify sources based on their identifiers in Gaia DR2. }

{At this point, we now have two lists: a subset of the nearest and brightest objects in the TIC and all nearby Gaia DR3 objects. We take the union of these two target lists, joining on Gaia DR3 IDs, and removing objects with duplicate IDs.
Joining the TESS and Gaia lists for nearby and bright objects yields several tens of thousands of sources, many of which are very faint and poorly characterized.
In order to narrow this list of objects down to the sources that will make good candidates for space-based direct imaging, we apply the magnitude cut discussed in Section \ref{considerations_section} to only include objects with TESS $T$ and Gaia $G$ magnitudes less than 12.} 

{Applying these cuts to the target list reduces our target list to a manageable size and restricts it to the population of nearby bright stars, of which the majority have more observations and are more well characterized than the average field star. 
For our combined list of nearby, bright TIC and Gaia DR3 objects, we gather measurements for their observable properties and estimates for their derived stellar properties. For every object with a Gaia DR3 identifier, we query the Gaia Archive to obtain the full table for Gaia DR3 data and astrophysical properties calculated via the Gaia Apsis pipeline\citep{gaiadr3,creevey2022,fouesneau2022}. We similarly gather the full TESS Input Catalog 8.2 entries for every object in our list with a TIC ID \citep{paegert2021}. Using this information, we further refine our target list by removing stars that the TIC identifies as artefacts or duplicate sources with other TIC objects, as well as removing objects that are tagged as galaxies.
We then query the CDS Simbad database for each object based on their primary identifier in their source catalog, namely their TESS or Gaia DR3 ID  \citep{wenger2000}. For the subset of objects where the identifier is not found by Simbad, we perform a second query using an alternative identifier such as Hipparcos, 2MASS, Tycho or Gaia DR2 IDs.
From Simbad we obtain additional identifiers not in the TIC as well as photometry, spectral types, spectroscopic properties, and the binarity of objects.  
In addition to the catalogs listed previously, we also use VizieR to gather several other tables for stellar measurements and derived properties \citep{vizier}. }

\subsection{Obtaining Stellar Properties}
For every star in the HPIC we aim to obtain maximum completeness in the astrophysical properties necessary for yield calculations, while also ensuring the {reliability} and self-consistency of properties. Measurements of these properties are often available from multiple sources, each with their own systematic uncertainties. In this section we describe our procedure for prioritizing  different sources of estimates and observations for each property.

\subsubsection{Positions and Proper Motions}
We start by compiling the positions of objects in the HPIC at epoch J2000, and obtaining their proper motions. Proper motions are useful for planing a direct imaging survey, because we would like to know the precise positions of where the stars we want to observe will be during the 2040s, when HWO is expected to launch. We obtain proper motions for each object via Gaia DR3 if available. Otherwise we use the measurements for proper motion given by the TIC 8.2. If an object doesn't have proper motion listed in Gaia DR3 or the TIC we obtain its proper motion from its Simbad entry. For a small subset of objects (74 stars in our final catalog), accurate proper motions {are not} available as they may be binaries or otherwise exhibit unusual nonlinear proper motions. 

\subsubsection{Photometry}

Measurements of a star's photometry will be important for a wide variety of use cases such as determining the required exposure times to detect planets, fitting the star's spectral energy distribution, and inferring its astrophysical properties via empirical relations. For each source in the HPIC, we obtain photometry in a wide variety of measured bands. The HPIC includes photometric measurements in the Johnson-Cousins \textit{UBVRI} bands, the 2MASS $JHK_s$ bands, {the TESS $T$ band,} and the Gaia $G$, $G_{BP}$, and $G_{RP}$ bands. We acknowledge that most objects will not have photometry in all of these bands, but we seek to gather all the measurements that are available. 

To obtain photometry for each star, we first check to see what information is available in the TIC. 
{All objects in the TIC have estimated magnitudes in the TESS $T$ band. For objects in our input catalog without $T$ magnitudes, namely those exclusively in the Gaia target list, we will calculate estimated $T$ magnitudes using the relations in \citep{stassun2019}, which compute TESS magnitudes as a function of Gaia band magnitudes. }
The TIC {also provides measurements and estimates of} $B$, $V$, $J$, $H$, $K_s$, and Gaia $G$ band magnitudes for many objects, so if the star in our catalog has some or all of these bands listed in the TIC, we use those values as a preferred source for the object's photometry. Next we check if the object is present in Gaia DR3. If so we obtain its $G$, $G_{BP}$, and $G_{RP}$ photometry from Gaia DR3, potentially overwriting any values in the TIC given by Gaia DR2. To supplement the Gaia photometry, we use the Gaia Catalog of Nearby Stars to see if corresponding 2MASS $JHK_s$ photometry is available \citep{GCNS}. 
Next, for each object we retrieve any missing photometry from Simbad if available. Finally, for stars still lacking photometry, we use photometry from the 10 Parsec Sample of \citet{Reyle2021} if the source is included in that catalog.

\subsubsection{Binarity and Contaminants}

One major consideration in determining the quality of a star as a direct imaging target is whether it is located close to another bright object. While the telescope's coronagraph can mask much of the light of a potential host star, light from nearby stars can bleed into the images, greatly increasing the exposure times required to find and characterize planets. The presence of a binary companion, or other bright contaminant sources at small angular separations from a host star, can therefore determine whether a star is a suitable target for direct imaging. In the HPIC, we determine whether each star is part of a known binary or whether it is close to a potential contaminant. We do not apply cuts to the HPIC based on binarity, but rather compile binary information which will allow for cuts to be applied later based on exposure time estimates. While obtaining fully characterized orbits for stars in binary systems would be ideal, for almost all binaries in our catalog this information is not available. For binaries and close neighboring stars we therefore compile their angular separations and magnitude differences.

To determine the binarity of a given star in the HPIC, we first check if the star has a Simbad entry and if Simbad provides a WDS ID for the object. For stars with a WDS ID, we retrieve the most recent measurements of their angular separation and magnitude difference (usually $V$ band) from the Washington Double Star Catalog \citep{wds_paper,wds_2023}. To find additional binaries in our catalog, we consult the list of multiple systems from the Gaia Catalog of Nearby Stars \citep{GCNS}. This catalog provides the Gaia IDs of companions, their angular separations as well as their magnitude differences in the Gaia $G$ band. 
While the Gaia mission has identified many other objects that are likely astrometric or spectroscopic binaries, they lack sufficient information to accurately assess how much of an obstacle they pose for direct imaging. It may be possible that some of these objects exhibiting nonlinear proper motions may be giant planet hosts, as the Gaia team has yet to release their full list of astrometric planet candidates. In order to not dismiss any potential targets of scientific interest out of hand, we keep these stars in our catalog, but add a flag to note that they were tagged as binaries by Gaia.

In addition to binaries, stars with a small angular separation from a target star will also serve to provide stray light and can adversely affect direct imaging observations. Therefore we would like to determine whether each object has any close neighbors that could be potential contaminants. For each object in the HPIC, we use the CDS X-Match service to identify all Gaia DR3 sources within 10 arcseconds \citep{boch2012}. We match all objects with J2016 positions to the J2016 positions of Gaia DR3 objects, and, for the few hundred objects without J2016 positions, we match their J2000 positions to those of Gaia objects. 
We record the Gaia IDs, angular distance, and $G$ magnitude difference of these objects in a supplemental table {hosted} alongside the HPIC.

\startlongtable
\begin{deluxetable}{l|l|l}
\tablecaption{Sources of stellar properties in the HPIC. Entries are ordered by their precedence in our pipeline. \label{APsources}}
\tablehead{
\colhead{Property}   & \colhead{Source} & \colhead{Number of Objects}
}
\startdata
    \multirow{8}{2cm}{Distance} & Total & 12801 \\ 
       ~ & 1. \citet{bailerjones2021} Gaia Distance & 12163 \\
       ~ & 2. TIC & 341 \\
       ~ & 3. Simbad & 136 \\
       ~ & 4. From Gaia parallax & 2 \\
       ~ & 5. From parallax in TIC & 70 \\
       ~ & 6. From parallax from Simbad  & 80 \\
       ~ & 7. Distance of known companion & 9 \\ \hline
    \multirow{10}{2cm}{$\mathrm{T_{eff}}$} & Total & 12782 \\
        ~ & 1. \citet{soubiran2022} & 2710 \\
        ~ & 2. TIC & 8852 \\
        ~ & 3. Gaia GSP-Spec & 550 \\
        ~ & 4. Gaia GSP-Phot & 58 \\
        ~ & 5. \citet{casagrande2011} & 16 \\
        ~ & 6. Simbad & 252 \\
        ~ & 7. \citet{mcdonald2017} & 32 \\
        ~ & 8. \citet{stassun2019} $\mathrm{T_{eff}}$ relation & 239 \\
        ~ & 9. \citet{pecaut2013} & 73 \\
        \tablebreak
       \multirow{9}{2cm}{Luminosity} & Total & 12684 \\ 
        ~ & 1. TIC Luminosity & 10813 \\
        ~ & 2. Gaia DR3 Flame & 501 \\
        ~ & 3. From \citet{casagrande2011} bolometric flux & 188 \\
        ~ & 4. \citet{stock2018} & 47 \\
        ~ & 5. Compute using Gaia bolometric correction & 932 \\ 
        ~ & 6. \citet{mcdonald2017}, small unc & 63 \\
        ~ & 7. \citet{pecaut2013} & 128 \\ 
        ~ & 8. \citet{mcdonald2017}, any value & 12 \\ \hline
        \multirow{2}{2cm}{Radius} & Total & 12668 \\ 
        ~ & 1. R from L and $\mathrm{T_{eff}}$ using Stefan-Boltzmann eq & 12668 \\ \hline
        \multirow{7}{2cm}{[Fe/H]} & Total & 11950 \\ 
        ~ & 1. See if available from same source as $\mathrm{T_{eff}}$ & 4028 \\
        ~ & 2. TIC & 7 \\
        ~ & 3. \citet{casagrande2011} & 1180 \\ 
        ~ & 4. Simbad & 2233 \\
        ~ & 5. GSP-Spec & 4144 \\
        ~ & 6. GSP-Phot & 358 \\ \hline
        \multirow{9}{2cm}{Mass} & Total & 12697 \\
        ~ & 1. \citet{mann2019} if M dwarf & 4388 \\
        ~ & 2. GAIA DR3 Flame & 4409 \\
        ~ & 3. TIC & 2707 \\
        ~ & 4. \citet{casagrande2011} & 250 \\
        ~ & 5. \citet{stock2018} & 47 \\ 
        ~ & 6. \citet{Jimenez2023} & 6 \\ 
        ~ & 7. \citet{kordopatis2023} & 189 \\ 
        ~ & 8. TIC $\mathrm{T_{eff}}$-Mass relation & 701 \\ \hline
        \multirow{2}{2cm}{Age} & Total & 4320 \\
        ~ & Same source as mass & 4320 \\ \hline
        \multirow{2}{2cm}{log(g)} & Either: Calculate using M and R & 12668 \\ 
        ~ & Or: Use spectroscopic value from same source as $\mathrm{T_{eff}}$ & 3350 \\
        \hline
\enddata
\end{deluxetable}

\subsubsection{Distances}
\label{distance_section}

The distance to a star plays an important role in determining whether a star is a good target in a direct imaging survey. As mentioned earlier, distance sets the angular scale of the habitable zone and provides a star's absolute magnitude, helping to constrain its luminosity. We therefore aim to provide the most accurate distances available for all of the stars in the HPIC. We used preliminary distance measurements in the selection of many of the target stars in our catalog, but we now gather the best distance measurements beyond what is available in a given {object's} source catalog. While ideally we would select the best measurements for distances and other stellar properties by determining which ones have the smallest uncertainties, in practice uncertainties in these measurements are often underestimated, unreported or subject to unaddressed systematic errors. We therefore chose to prioritize measurements based on their source catalog rather than relying on individual uncertainty estimates.

We obtain the distances for all stars in the HPIC prioritized from sources in the order shown in Table \ref{APsources}. First we determine if the object has a Bayesian estimate of its Gaia DR3 distance calculated by \citet{bailerjones2021}, and use that for the object's distance if available. As indicated in the first section of Table \ref{APsources}, the majority of objects in the HPIC will use these Gaia DR3 distances. For the subset of objects that don't have accurate Gaia distances, either because they are too bright or are otherwise unable to have precise parallax measurements from Gaia, we consult other catalogs. First we identify objects with distances listed in the TIC. The TIC gathers distance measurements from a variety of sources, most notably Hipparcos and Gaia DR2, and we use those measurements when available. A few objects in the TIC have distances listed, but do not provide a source for the measurement or estimates for the uncertainty. In that case, we shall prefer other sources of distance measurements unless no others are available. If an object does not have a reliable distance from the TIC, we use the distance provided by Simbad if available. Since Simbad distances are from a variety of sources of varying accuracy, we require that the distance given is roughly consistent to that derived from one over its parallax (to within $20\%$).
If an object does not have a distance measurement from any of the preceding sources we calculate its distance using one over its parallax if available. In order of precedence we prefer to use parallax measurements from Gaia DR3, then those provided in the TIC, then those from Simbad. Finally, a small number of remaining  objects in binaries may not have well constrained distances, but their companions may. In that scenario{,} we use the distance to the companion as the object's distance for lack of a better measurement. 

\subsubsection{Effective Temperature}

Stellar effective temperatures, while not directly used in yield calculations, are very important for our understanding of the spectral types and estimated colors of stars, and can be used to calculate properties that are critical to exoplanet yield calculations, such as luminosities. To obtain measurements or reasonable estimates of $\mathrm{T_{eff}}$ for each star we follow the procedure shown in Table \ref{APsources}.  For every object in our target list, we first identify if it is included in the Pastel catalog of \citet{soubiran2022}. If it is, we use that value for the effective temperature, otherwise we check to see if the object has an effective temperature given in the TIC. We use the $\mathrm{T_{eff}}$ measurements provided in the TIC when available. For a subset of objects in the TIC, effective temperatures were calculated using Gaia photometry via an empirical relation provided in Table 2 of \citet{stassun2019}. If newer Gaia DR3 photometry is available for these objects, we use it to update their $\mathrm{T_{eff}}$ estimates.

If an object is not in the TIC or doesn't have an effective temperature in the TIC, we look up its Gaia DR3 ID and determine whether its effective temperature has been calculated by the Gaia mission. If the object has an effective temperature calculated from the Gaia mission's General Stellar Parametrizer from Spectroscopy (GSP-Spec), we use that value. Otherwise, {we use estimates of an object's } $\mathrm{T_{eff}}$ calculated using Gaia's General Stellar Parametrizer from Photometry (GSP-Phot) pipeline {when available}. 
If an object does not have Gaia estimates of $\mathrm{T_{eff}}$ we check if the object's HIP ID is present in the \citet{casagrande2011} reanalysis of the Geneva-Copenhagen Survey (hereafter referred to as the GCS reanalysis), and use that value if it is available. Next we check if the object has a Simbad entry and use the value for effective temperature that it provides. If Simbad does not provide $\mathrm{T_{eff}}$ for the object, we see if it is in the catalog of \citet{mcdonald2017}, containing the stellar properties of Tycho-2 and Hipparcos stars derived using Gaia DR1 distances, and use their $\mathrm{T_{eff}}$ when available.

For the remaining stars without $\mathrm{T_{eff}}$ measurements, we use empirical relations to estimate their effective temperatures. First we use the empirical relation from TESS described earlier, which calculates the $\mathrm{T_{eff}}$ using Gaia colors \citep{stassun2019}. If objects have Gaia photometry and $G_{BP} -G_{RP}$ colors between -0.2 and 3.5, we use this empirical relation to calculate the effective temperature. For stars that fall outside the range of applicability for the previous relation, we use the relation to \citet{pecaut2013} (see their Table 5) to obtain $\mathrm{T_{eff}}$ from $B-V$ colors. If a star in our catalog is a known dwarf star and has measured $B$ and $V$ photometry we can use this relation to obtain $\mathrm{T_{eff}}$.

\subsubsection{Luminosity}


To calculate the luminosities of objects in the HPIC we use the procedure shown in Table \ref{APsources} and described below. 
For all objects in our target list, we first check if the object is in the TIC and if it has a luminosity provided. For objects with TIC luminosities, we check if the luminosity was calculated using Gaia DR2 measurements. If this is the case we check if newer Gaia DR3 photometry is available. If the object has newer DR3 photometry, we calculate an updated TIC luminosity using the Gaia bolometric correction as a function of $\mathrm{T_{eff}}$ provided in \citet{stassun2019}. If an object’s luminosity was not calculated using Gaia photometry, or if DR3 measurements are not available, we instead use the luminosity that is given in the TIC{, updating it when revised distance estimates are available}.

If an object doesn't have a luminosity in the TIC, we check to see if it has a corresponding Gaia DR3 ID. If an object in Gaia DR3 has a luminosity calculated by the Gaia FLAME pipeline, we use that value for L. For objects without TIC or Gaia luminosities we determine whether the object is present in the GCS reanalysis of \citet{casagrande2011}. If the GCS reanalysis provides a bolometric flux for the object, we use the known distance to calculate its luminosity. Next, if an object is a giant star, we use the \citet{stock2018} list of stellar parameter for giant stars to obtain the object's luminosity if available.

For objects that still don't have luminosities, we determine if it has Gaia DR3 photometry and falls within the valid temperature range of the bolometric correction for main sequence stars. If so, we calculate its luminosity using the TIC relations between G band magnitudes, effective temperatures, and luminosities \citep{stassun2019}. The next step in our procedure is to determine whether the object has a luminosity estimate in the catalog of \citet{mcdonald2017}. We should note that while most luminosities in this catalog are accurate, there are a few stars with very highly uncertain luminosities. We choose to only use stars with less than $30\%$ uncertainty in luminosity from this source, unless no other sources of luminosity values as available. Finally, if an object is a dwarf and has photometry in the B and V bands we use the bolometric correction of \citet{pecaut2013} along with the object's distance to calculate L.

\subsubsection{Radius}

For stellar radii of objects in the HPIC we aim for self-consistency with the other measurements listed. {While independent measurements and estimates of the radii of stars are often available in the literature, for this catalog we opt to calculate stellar radius using luminosity and effective temperature in the Stefan-Boltzmann equation. This ensures that the radius we provide will be consistent with the values obtained for L and $\mathrm{T_{eff}}$ in the previous sections.} 


\subsubsection{Metallicity}

Metallicities, while not required in yield calculations, provide fundamental information about composition of a star and can be used to better constrain the star's mass, age and evolutionary track. To obtain metallicities for a star in the HPIC, we first see if [Fe/H] is available from the same source as the effective temperature. This is to ensure consistency between the measurements that we choose to use. If no metallicity is available from the source of $\mathrm{T_{eff}}$, we then check if the star has an entry in the Pastel catalog and use its [Fe/H] value when present \citep{soubiran2022}. Next we determine if the object metallicity provided in the TIC, or if not, whether the object has a metallicity in the GCS reanalysis \citep{casagrande2011}. 

For objects still without metallicities, we look up the star's Simbad entry and obtain it's [Fe/H] value when available. Finally, if a star lacks a metalliticity after the previous steps, we check to see if it has a value calculated from the Gaia mission's automated pipelines. We identify if there is an [Fe/H] value from the GSP-Spec pipeline, derived from spectroscopic measurements, or if not whether there is a photometric estimate from the GSP-Phot pipeline.

\subsubsection{Mass}

Masses of stars in our catalog will be important, not only for determining the orbits and masses of potential planets, but also for understanding the evolution and properties of the host star. To calculate the mass of a star in the HPIC we first identify if it is a cool dwarf star. If its effective temperature is less than 4000K, we use the relation of \citet{mann2019} to obtain the masses of M and late K stars from their absolute 2MASS $K_S$ magnitudes. For {a star} outside of this temperature range or without {a $K_S$ magnitude}, we determine whether it has a mass estimate provided by the Gaia FLAME pipeline and use it if available. Alternatively if the object is in the TIC 8.2 and has a mass estimate provided, we use that value.
Next we investigate whether the object has a mass available in the GCS reanalysis of \citet{casagrande2011}. If the object is a giant star, we check to see whether its mass has been calculated by \citet{stock2018}. Alternatively if it is a white dwarf, we obtain its mass from the \citet{Jimenez2023} catalog of white dwarfs within 100 pc.
For objects that still lack masses, we consult the \citet{kordopatis2023} isochrone fits for Gaia DR3 stars an see if the object in question has an age estimate. Finally for the remaining objects, we estimate their masses using the empirical relation between $\mathrm{T_{eff}}$ and mass from \citet{stassun2019}.

\subsubsection{Age}
\label{age_section}
Ages are among the most difficult stellar properties to accurately constrain \citep{Soderblom2010}. While we ideally would like to obtain ages for all the targets in our catalog, it is only feasible to do so for the most well characterized stars. For the purposes of the HPIC, we obtain stellar ages when available from the same sources as the star's mass. In planning for HWO, more work is needed to determine the ages of the best direct imaging target stars. Future studies using data from the TESS and upcoming PLATO missions will be able to obtain measurements of stellar rotational periods for use in gyrochronology and asteroseismic measurements to better constrain stellar ages \citep{TESS_overview,PLATO_overview}.

\subsubsection{Surface Gravity}
\label{logg_section}
{Stellar surface gravities provide important information relating to a star's luminosity class (i.e. if it is a dwarf or a giant), and they are useful for stellar modelling, allowing additional stellar properties to be matched beyond a star's location on the HR diagram. While log(g) is not required for yield calculations, it will be important for understanding the stellar population that HWO plans to survey.} 

{We aim to provide self-consistent log(g) values for all stars in our sample, but in doing so we are faced with the question as to what we define to be self-consistent. There are two potential ways to be self-consistent in log(g): one could calculate log(g) using the stellar bulk properties of mass and radius or one could obtain log(g) spectroscopically from the same source as other spectroscopic properties in our catalog such as $\mathrm{T_{eff}}$ and [Fe/H]. In the HPIC we include log(g) from bulk properties and spectroscopic log(g) as separate quantities in order to not make a judgement as to which form of self-consistency is of greater importance. }

{For every star in the HPIC with a known mass and radius, we calculate its bulk log(g) using the relation $\log(g) = \log(g_{\odot}) + \log(M/M_{\odot}) - 2\log(R/R_{\odot})$. We obtain spectroscopic log(g) as a separate quantity, identifying if it available from the same source as the spectroscopic measurements of $\mathrm{T_{eff}}$. Note that for spectroscopic log(g) measurements, we opt not to use values from Simbad, as a few reported literature values of log(g) from Simbad can be wildly inaccurate, and the values of log(g) that Simbad reports come from a wide variety of sources, with not all estimates guaranteed to be from spectroscopy rather than calculated values from bulk properties. In practice, of the stars in our catalog using $\mathrm{T_{eff}}$ from Simbad, very few ($\sim$18) have log(g) measurements available from the same source, so excluding these measurements doesn't have a significant effect on catalog completeness.} 

{Most stars in our catalog will have bulk log(g) estimates, but only a subset will have reliable spectroscopic log(g). As we report two separate values for log(g) in the HPIC, we recommend the following procedure when determining the ideal log(g) value to use for a given star: Use spectroscopic log(g) when available, and if not, use log(g) calculated using stellar bulk properties.} 

\begin{deluxetable}{l|l|l}
\tablecaption{Completeness of astrophysical properties and photometry in the HPIC.   Total number of objects: 12944. \label{completeness_table}}
\tablehead{
\colhead{Property}   & \colhead{Number of Objects} & \colhead{Completeness (\%)}
}
\startdata
\sidehead{Astrophysical Properties} \hline
Distance   & 12801                   & 98.9             \\ 
$\mathrm{T_{eff}}$       & 12782                   & 98.7              \\ 
L          & 12684                  & 98.0             \\ 
Radius     & {12668}                  & {97.9}             \\ 
$\mathrm{[Fe/H]}$ & 11950                  & 92.3             \\ 
log(g)     & {12711}                  & {98.2}             \\ 
Mass       & 12697                  & 98.1             \\ 
{Age}        &  {4320}    & {33.4} \\ \hline
\sidehead{Photometry} \hline
$U$          & 1397                  & 10.8             \\ 
$B$          & 12430                  & 96.0             \\ 
$V$          & 12672                  & 97.9             \\ 
$R$          & 7595                  & 58.7             \\ 
$I$          & 2220                  & 17.2             \\ 
$J$          & 12614                  & 97.5             \\ 
$H$          & 12610                  & 97.4             \\ 
$K_s$          & 12605                  & 97.4             \\ 
Gaia $G$     & 12580                  &  97.2  \\
{TESS $T$ (calculated)}     & {12944}     & {100.0}
\enddata

\tablecomments{{The value of log(g) completeness in this table includes all objects that have either a bulk log(g) estimate or a spectroscopic log(g) measurement.}}
\end{deluxetable}

\begin{figure}
    \centering
    \plotone{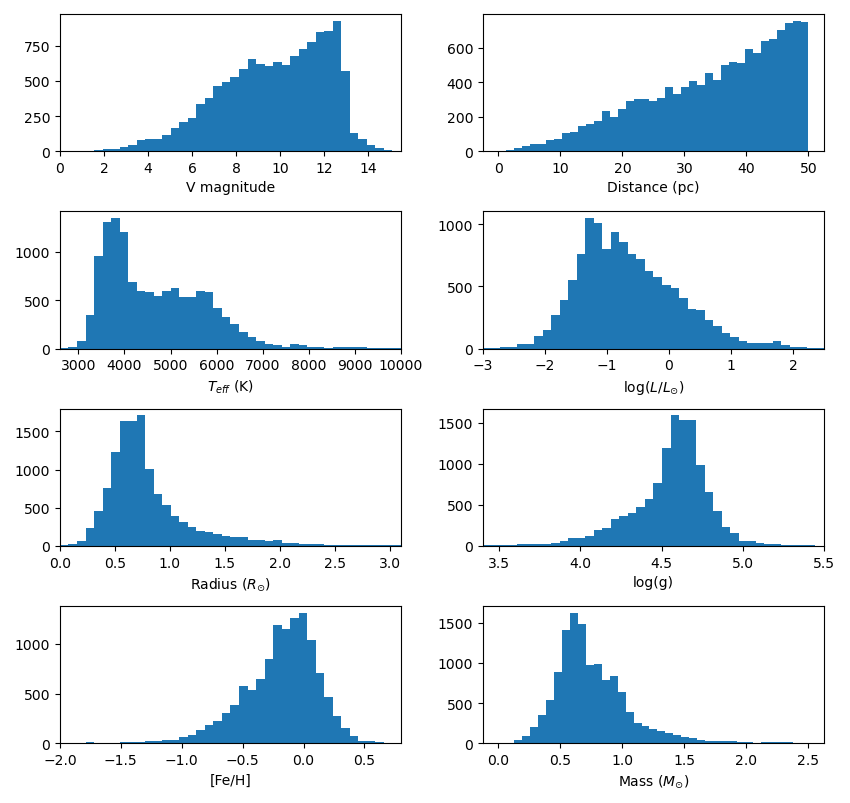}
    \caption{Histogram of stellar properties for stars in the HPIC. Note that a few outlier stars fall outside the axis range of some of the histograms. {Surface gravity shown here is log(g) calculated using stellar bulk properties.}}
    \label{combined_hist}
\end{figure}

\subsection{ {Completeness of Stellar Properties}}
The results of our automated pipeline to gather and compute stellar properties for stars in the HPIC are shown in Table \ref{completeness_table} and in the histograms shown in Figure \ref{combined_hist}. The HPIC contains 12944 objects, and for each of these objects we aim for the highest possible {catalog} completeness in astrophysical properties and photometry.  {Note that in this section we are referring to ``catalog completeness'' in terms of the fraction of stars in our catalog with a given property rather than ``survey completeness'' which we will discuss in Section \ref{yield_section}}. We can see that for most astrophysical properties {(excluding ages)} the {catalog} completeness is on the order of $98\%$ or higher. Many of the objects lacking fundamental properties fall in a few main categories. There are binaries which don't have accurate parallax measurements, newly resolved binary companions which have not been studied in detail, or distant field objects lacking distance measurements (thus causing them to remain in our target list when they should have been removed). Excluding this $2\%$ of objects from our target list is unlikely to have noticeable effects on yield calculations, but we choose to keep these objects so that their properties can be updated with the results of future studies prior to HWO. {Of the various astrophysical properties that we gather in our pipeline, it is unsurprising that stellar age has much lower catalog completeness due to the difficulty in obtaining age estimates (see Section \ref{age_section}). Other than ages, [Fe/H] measurements have slightly lower catalog completeness than the other astrophysical properties listed.} 
This is because determining [Fe/H] typically requires analyzes of host star spectra, which may not be available for lesser studied target stars.

In terms of photometry, all stars in our catalog have photometry in multiple bands, but they differ in which bands are available. Most objects do not have photometry in all of the bands we gathered measurements for, but some bands such as Johnson $B$ and $V$, 2MASS $JHK_s$, and Gaia $G$ bands have close to $100\%$ completeness. {The TESS $T$ band has $100\%$ completeness in our catalog, but these values are calculated estimates rather than measurements.}  Other bands, namely the $U$,$R$, and $I$ bands are only available for a limited sample of stars in our catalog due the fact that the stars either haven't been surveyed in these bands, or, in the case of the U band, the objects were too faint to be detected at the given wavelength.

\begin{figure}
    \centering
    \plotone{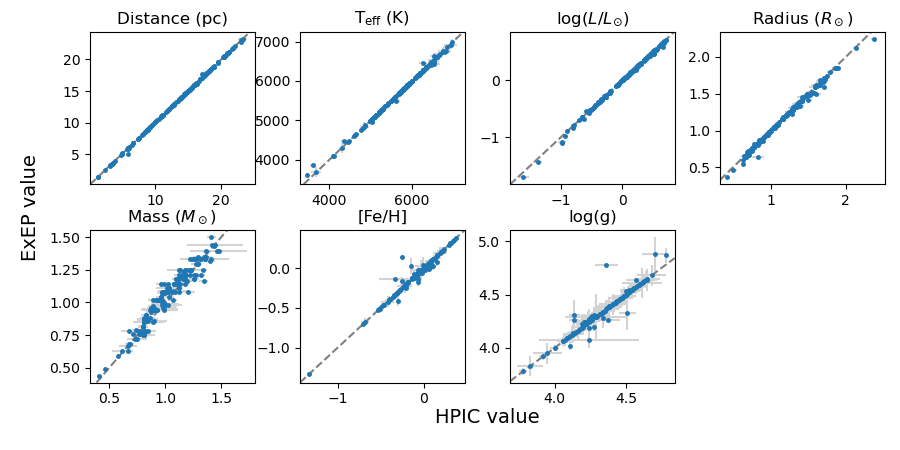}
    \caption{Comparison of stellar properties in the HPIC (x axes) and ExEP HWO precursor science list (y axes). Uncertainties are plotted when available, though note that for a few properties and individual stars uncertainties are not available from both catalogs. Stellar properties are consistent between these two catalogs, excluding one or two outliers.}
    \label{benchmark_fig}
\end{figure}

\begin{figure}
    \centering
    \plotone{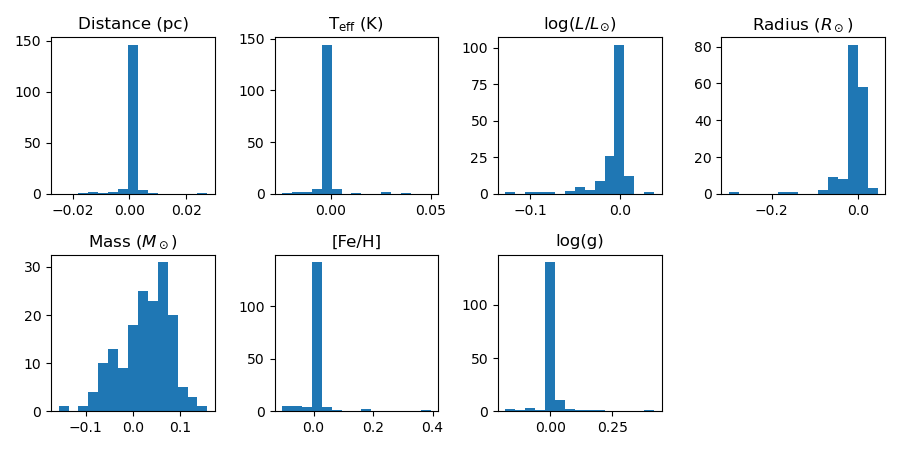}
    \caption{Residuals between HPIC and ExEP values of stellar properties. Note that fractional residuals are shown for all quantities except log(L), [Fe/H], and log(g) where differences are shown. Single outliers have been omitted in a few subplots in order to make the distribution of residuals more clear to see.}
    \label{residuals_fig}
\end{figure}

\subsection{{Consistency} of Stellar Properties}

Now we assess the degree to which our automated pipeline provides values for each of the stellar measurements {which are reliable, precise and consistent with other literature source}. In order to do this, we compare the properties gathered and calculated in the HPIC to those of a trusted source that was not directly drawn from in the creation of our target list. 
We adopt the NASA ExEP Mission Star List for the Habitable Worlds Observatory  as our baseline for comparison \citep{mamajek2023}.
As mentioned earlier, this list of the 164 best targets for space-based direct imaging is curated by hand and contains the most heavily vetted measurements for the properties of these stars. 

Figure \ref{benchmark_fig}, shows the comparison of the values of astrophysical properties between the two target lists. All of the stars in the ExEP list are contained within the HPIC, and are shown in this figure. {As we have two separate estimates for log(g) in our catalog, we used the procedure described in Section \ref{logg_section} to obtain a best estimate of log(g) on a per star basis in order to compare to the values in the ExEP list. We use spectroscopic log(g) values when available and otherwise use bulk values.}
In Figure \ref{residuals_fig} we plot the fractional residuals between the two lists. Note that for the values of log(L), [Fe/H], and log(g) we plot the differences rather than fractional differences to avoid zeros in the denominator.  Looking at these plots we can see that, for the most part, the values for each parameter adhere closely to a one to one correlation between both target lists. For some properties there is a clustering of residuals around zero{,} as in some cases the same source was used to gather the value of an astrophysical property in both the HPIC and ExEP lists. This is most apparent for distances, as both catalogs primarily use distances from Gaia DR3 parallaxes. We notice that for a few lower mass stars there tends to be a small discrepancy in $\mathrm{T_{eff}}$ and log(L), likely caused by the fact that these stars may not be as well characterized or may be more difficult to characterize than other stars, resulting in greater variation in measured properties between sources. This discrepancy propagates to our values of stellar radii, which for lower radius stars are slightly smaller in the HPIC than in the ExEP list. Still, accounting for the uncertainties in radius values, {most} of these discrepant values are still consistent with the ExEP value within 1 - 2 sigma.

There is a larger spread in stellar mass values between the two catalogs, but such is to be expected as stellar masses tend to be more difficult to constrain than other stellar properties so one would expect larger variability between different estimates. The values for stellar masses are consistent to within around $10\%$, which is notable as the uncertainties on individual estimates often exceed that. $10\%$ precision in mass is more than sufficiently precise for use in yield calculations, as the primary use of stellar masses is to derive planetary orbital periods which go as $M^{-0.5}$ for a given semimajor axis and are thus less sensitive to uncertainties in mass. Mass estimates in the ExEP list tend to be slightly higher than those in the HPIC, likely representing systematic differences in the means masses were acquired. 
The values of [Fe/H] and log(g) appear to exhibit larger scatter from a one to one correlation in Figure \ref{benchmark_fig}. However, given the larger uncertainties in these properties, the values in both lists remain consistent with each other except for the case of one or two outliers.

\subsection{Contents of the HPIC}

The HPIC contains the stellar properties necessary for exoplanet yield calculations, as well as additional parameters that are useful to characterize exoplanet host stars. A list of the columns included in the HPIC can be found in Appendix \ref{HPIC_columns}, while a full list of the source and quality flags can be found in Appendix \ref{HPIC_flags}. {The HPIC is publicly available and hosted by the NASA Exoplanet Archive: \dataset[doi:10.26133/NEA39]{https://doi.org/10.26133/NEA39} \citep{HPIC_dataset}. }

The HPIC is not the final input catalog for the Habitable Worlds Observatory, but rather it serves as the most complete list so far of potential targets for space-base direct imaging with HWO. The HPIC represents the state of our current knowledge about the stellar properties of the population of bright and nearby stars. However, our knowledge of stellar properties is constantly evolving, and in advance of HWO's launch in the 2040s we expect to learn more about these target stars. Current missions like TESS and Gaia, as well as upcoming missions such as PLATO, will help us to better constrain the stellar properties of these stars and identify new planet hosts \citep{TESS_overview,prusti2016,PLATO_overview}.
As our knowledge of these stars is continually improving we intend to keep the HPIC regularly updated with the results of new studies such as future Gaia data releases. In future updates of the HPIC we would like to add measurements of additional astrophysical properties, such as those which NASA's ExoPAG SAG 22 identified as being most useful to the astronomical community \citep{SAG22_report}. For example we like to obtain stellar properties including stellar activity indicators, disk properties, full binary orbits, and X-ray and UV fluxes. Additionally, one of the most useful but also difficult to obtain stellar properties is the age of a star. In the HPIC, we currently have stellar ages listed when obtained from the same source as the stellar mass, but in future updates we would like to increase our completeness in stellar ages. This would involve gathering age estimates from a wide variety of sources, such as ages obtained via asteroseismology or gyrochronology. 
While it won't be feasible to obtain these different stellar property measurements for all stars in our sample, in future releases of the HPIC we would like to determine which stars have measurements of certain less commonly obtained properties.

\section{Yield calculations}
\label{yield_section}

We now demonstrate the utility of the HPIC {for} performing exoplanet yield calculations. We do not aim to accurately predict the yield for any specific mission design here, and will therefore not focus on specific mission parameter details. Rather, our aim is to demonstrate how improving the mission design or varying the science goals may require a stellar catalog as expansive as the HPIC.

\begin{deluxetable}{l|l|l|l|l|l}
\tablecaption{Details for different yield calculations \label{yield_table}}
\tablehead{\colhead{Name} & \colhead{Telescope} & \colhead{Planet Type}&\colhead{Survey Duration} & \colhead{Total Yield} & \colhead{Number of targets}}
\startdata
        Baseline & LUVOIR B & Exo-Earth & 2 years&  29.07 & 215 \\
        Updated Target list & LUVOIR B & Exo-Earth & 2 years& {28.48} & {211} \\
        Super LUVOIR B Survey & Super LUVOIR B & Exo-Earth & 2 years & {60.85} & {432}\\
        Cool Jupiter Survey & LUVOIR B & Cool Jupiters & 6 months& {218.98} & {1109} \\ \hline
\enddata  
\end{deluxetable}


\begin{figure}
    \centering
    \plotone{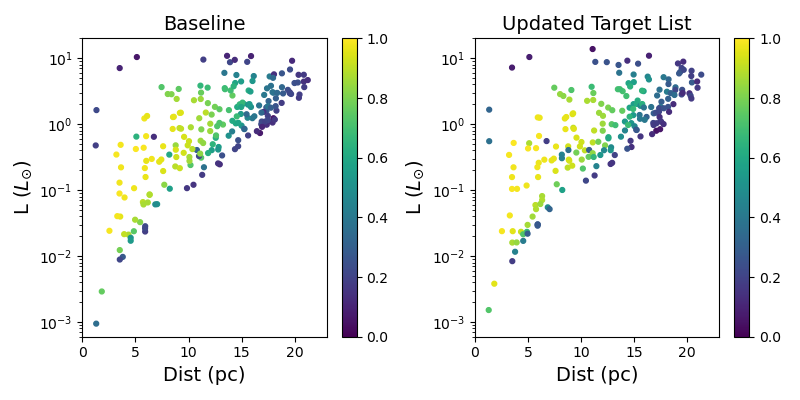}
    \caption{Comparison of the stars selected for a two year exoEarth candidate survey using the baseline AYO input catalog and the updated HPIC. Points are color coded by HZ completeness. Details about the different yield calculations are given in Table \ref{yield_table}. The population of stars selected does not change significantly when updating the target list, but the completeness of individual stars and their relative priority can change.}
    \label{yield_figure}
\end{figure}

\subsection{Effects of an improved stellar catalog\label{section:yield_improved_catalog}}

First we examine the impact of the HPIC on exoEarth candidate (EEC) yields. To do so, we perform two identical yield calculations, one using the AYO input catalog used by \citet{Stark2019} and one using the HPIC. For both calculations, we simulate a two year EEC survey using the LUVOIR-B-like scenario detailed in \cite{Stark2019}. Briefly, this consists of a $6.7$ m inscribed diameter telescope with parallel UV and visible wavelength coronagraph channels using deformable mirror-assisted vortex charge 6 coronagraphs. We make identical astrophysical, mission, and survey assumptions to those made in \citet{Stark2019}, with three exceptions. We increase the spectral resolution for characterization from $R=70$ to $R=140$ to be consistent with the recent results of \citet{latouf2023}. This change alone would lower the expected EEC yield by $\sim$10\%. However, we make two additional changes that recover this yield loss. First, because we are interested in how the target selection changes with the input stellar catalog, we adopt identical exozodi levels for all stars (three zodis) instead of randomly drawing from a distribution. Second, we include some stars from the original AYO catalog that were excluded from previous calculations. The original AYO input catalog of \citet{Stark2019} cut all stars without spectral classifications in an effort to avoid spurious sources. However, using the HPIC, we are able to verify that almost all of these stars are indeed valid sources. Therefore, to ensure a valid comparison, we allow AYO catalog entries without stellar classifications. 

To simulate the EEC survey, we distribute planets over the \citet{kopparapu2013} optimistic HZ, ranging from $0.95$ AU to $1.67$ AU for a solar twin. We distribute planet radii from $0.6$ to $1.4$ Earth radii with a stellar insolation-dependent lower limit, consistent with the EEC definition established by the LUVOIR and HabEx studies \citep{LUVOIR_final_report,HabEx_Final_report}. We adopt the occurrence rates of \citet{Dulz2020}, such that $\eta_{\oplus}=0.24$, and sample all possible orbits and phases. We use the AYO method detailed in \citet{Stark2019} to optimize target selection and exposure times in order to maximize the expected yield of the mission.

The results of our yield calculations with the original AYO input catalog and the HPIC are listed in Table \ref{yield_table}. For our baseline case using the original AYO input catalog, we estimate an EEC yield of{ 29.07 with 215} stars selected for observation. Using the updated HPIC, the yield decreases negligibly to {28.48 with 211} stars selected for observation. The population of stars selected for each survey is shown in Figure \ref{yield_figure}, color coded by HZ completeness. Among these objects are the best candidates for exoplanet direct imaging identified by the ExEP mission stars list.

{Eleven of the 211} stars selected from the HPIC are newly added stars that don't appear in the original AYO {output target list}. These newly added stars mostly serve to replace stars from the previous list which were no longer selected, mainly due to revisions in their stellar properties or newly discovered binary companions that cause stray light and increased exposure times. A few of the {211} stars selected from the HPIC were available in the original AYO stellar catalog but went unselected in our baseline calculation. These stars were selected either because new estimates of their properties made them better options or simply because they were filling the spots left behind by objects that decreased in {habitable zone} completeness. Many of the selected stars were the same between both calculations, but changed in expected {HZ} completeness due to revisions to their properties. We conclude that updating the quality and comprehensiveness of the target list does not have a significant impact on the total expected EEC yield, but does affect which individual stars are selected, as well as their exposure times and expected completeness.

\begin{figure}
    \centering
    \plotone{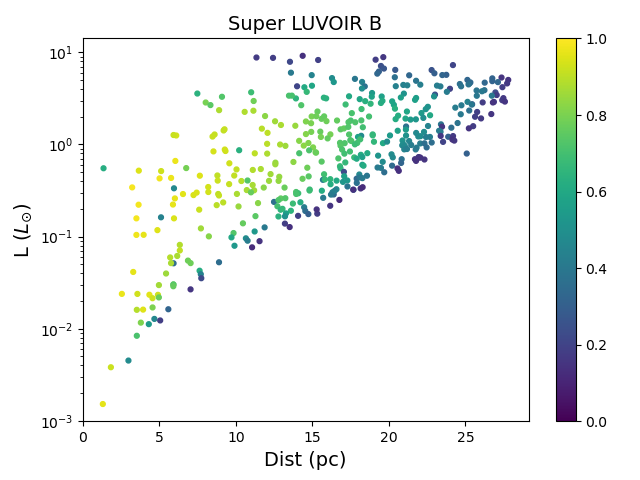}
    \caption{Stars selected for a two year exoEarth candidate survey using an $8$ m inscribed diameter "Super LUVOIR B" telescope. Points are color coded by HZ completeness. About twice as many stars as the baseline LUVOIR B case are selected, extending to larger distances.}
    \label{super_luvoir_fig}
\end{figure}

\subsection{Modifying the telescope design}
We do not yet know the design details of the HWO mission. One of the motivations for broadening the input catalog and extending the completeness and accuracy of stellar properties is to enable accurate yield calculations for a wide variety of telescope designs. Mission architectures that differ substantially from the HabEx and LUVOIR concepts could have significant differences in the population of stars they are able to observe. 

To demonstrate the utility of the HPIC for studying missions that deviate from the LUVOIR-B baseline, we consider a larger telescope with higher throughput than LUVOIR-B.  We refer to this hypothetical scenario as a ``Super LUVOIR-B''.  Specifically, we choose an inscribed diameter of 8 m (compared to the 6.7 m inscribed diameter of LUVOIR B) and increase the end-to-end throughput by a factor of four. This factor of four in throughput serves to represent plausible reductions in exposure times from tangible design trades, such as a reduction in aluminum reflections, parallelized dual visible wavelength channels, improved coronagraph design, and/or improved PSF calibration techniques. Using this telescope design, we again simulate the EEC yield using the HPIC.

Figure \ref{super_luvoir_fig} and Table \ref{yield_table} shows the results of the ``Super LUVOIR-B'' yield calculation. The EEC yield increases by more than a factor of two when compared to the baseline LUVOIR-B yield calculation. Importantly, this increase in yield comes in a large part from an expanded target list, {more than} twice the size of the baseline target list. Comparing the ``Updated Target List'' plot in Figure \ref{yield_figure} with Figure \ref{super_luvoir_fig}, one can see that the selected target list extends to more distant stars. 

A few objects from the ``Updated Target List'' case were not selected as targets in the ``Super LUVOIR-B'' case. These objects, mainly close-by stars with high stellar luminosities and low HZ completeness, appear to not have been included in the super LUVOIR B list of targets because better candidates at larger distances were made available by the larger telescope diameter and throughput. Since this survey has more choice in potential direct imaging targets, it doesn't need to settle for these lower completeness targets.

\begin{figure}
    \centering
    \plotone{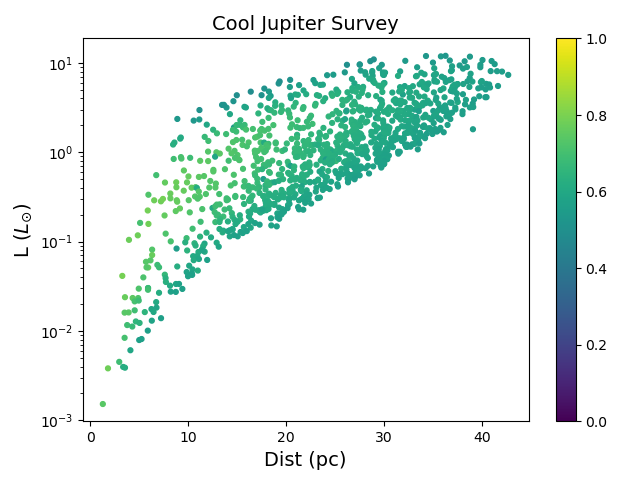}
    \caption{Stars selected for observation during a six month cool Jupiter detection survey. Points are color coded by {survey} completeness. Around 900 more stars were selected for observation that the baseline exo-Earth survey, many of which are fainter and at larger distances.}
    \label{cool_jup_fig}
\end{figure}

\subsection{Changing the science goal}

Studies of space-based direct imaging mission concepts have primarily focused on the detection of Earth-like planets \citep{brown2005,Savransky2010,Stark2014,morgan2019}, and for good reason -- this is a primary science motivator that drives mission design. However, this is not the only exoplanet science that could result from HWO; a mission designed to characterize potentially Earth-like planets could \emph{excel} at detecting Jupiter analogs. 

To demonstrate how the HPIC can be used for surveys that aren't solely focused on imaging Earth analogs, we simulate a six month survey to detect cool Jupiters using the LUVOIR-B-like design discussed in Section \ref{section:yield_improved_catalog}. For our ``cool Jupiters" we adopt planets ranging from $6.0$ to $14.3$ Earth radii. We use an inner semi-major axis of $1.54$ AU, consistent with the ``cool" temperature classification of \citet{Kopparapu2018}, but constrain the outer semi-major axis to $5$ AU, scaling the semi-major axis constraints with stellar insolation. We again adopt the \citet{Dulz2020} occurrence rates, resulting in $\eta_{\rm CJ}=0.32$. To illustrate the potential magnitude of a target list for an alternative science goal, we do not require spectral characterization for these observations and focus only on $V$ band detections.

Table \ref{yield_table} summarizes the results of our calculation and Figure \ref{cool_jup_fig} shows the population of stars selected for observation. The potential yield of this six month detection survey is {219} planets. The selected target list for this survey is much larger than that of the EEC survey, totaling more than {$1100$} stars and extending to fainter magnitudes and larger distances. Figure \ref{cool_jup_fig} shows that compared to the EEC survey, the target list shifts toward later type stars. While the HZs of these stars are within the coronagraphic IWA and thus are not good targets for the EEC survey, they are acceptable direct imaging targets for the more distant cool planets. We note that the minimum completeness in Figure \ref{cool_jup_fig} is $\sim50\%$. This therefore represents only a \emph{fraction} of the stars that are potential cool Jupiter target stars; extension of the survey duration beyond six months could substantially expand {the number of targets}.

\section{Conclusions} 

In this study we constructed the HPIC, a new input catalog of roughly 13,000 bright nearby target stars for space-based direct imaging using the Habitable Worlds Observatory (HWO). Because HWO is in the very early stages of planning and does not yet have a finalized mission architecture, this input catalog is agnostic of potential telescope and coronagraph designs.
To maintain flexibility and to easily update the HPIC in the future, we developed an automated pipeline to construct the input catalog and obtain measurements and estimates of stellar properties for each object. The HPIC obtains high completeness in stellar photometry, measured properties and derived fundamental properties, while also maintaining consistency between values.
We benchmarked the results of our catalog to the manually curated ExEP target list and showed that stellar properties are in good agreement for the highest priority HWO target stars.

The increased breadth of targets and {reliability} of stellar properties in the HPIC allowed us to accurately predict exoplanet yields from vastly different telescope designs and surveys focusing on different planet types. Using the HPIC as an input for an exo-Earth survey using a LUVOIR B design had a negligible effect on the net yield of the survey when compared to the results using an earlier target list, but there was a noticeable change in the population of stars selected to be surveyed and the priority given to specific stars. A survey using a telescope with a larger diameter and higher coronagraph throughput than LUVOIR B resulted in around twice the yield of exo-Earths compared to the LUVOIR B case and {more than} double the number of stars surveyed, extending to stars at farther distance from Earth. Similarly a shorter survey using LUVOIR B to observe cool Jupiters could observe around 900 more targets than an exo-Earth survey and would probe a population of stars that were more distant with higher apparent magnitudes, including more stars with later spectral types.
For these yield calculations, many of the targets selected for observation were not present in the baseline LUVOIR-B output target list. Critically, surveys using more ambitious telescope designs or those focused on different planet types often observed a population of fainter, more distant target stars. The HPIC is complete to fainter magnitudes than the \citet{Stark2019} target list, and contains more accurate measurements for their stellar properties, ensuring that we can properly model the yield{s} of these surveys.

{Our catalog provides measurements and derived properties for the population of nearby bright stars, and it will be useful for a wide variety of use cases in general astrophysics beyond precursor science for HWO. 
In particular it will benefit exoplanet surveys using methods other than direct imaging. Each exoplanet detection method has its own set of biases and detection requirements, meaning that they are sensitive to planets around different populations of host stars. NASA's ExoPAG SAG 22 identified the regions of stellar distance and luminosity space probed by different methods and found that the region with the highest overlap was for nearby FGK and early M stars \citep{SAG22_report}. The population of stars in the HPIC fall in this overlapping region of parameter space, and our work to constrain their stellar properties will allow other detection methods to more precisely estimate planetary properties and obtain a suite of properties unavailable using individual detection methods alone.
Surveying the stars in the HPIC using other exoplanet detection methods will also allow new planet candidates to be found in advance of HWO, and could influence HWO target selection based on precursor information.}

{As we lay the first steps in preparing for HWO, the HPIC will grant us a better understanding of the population of stars that will be surveyed, allowing future trade studies and yield calculations to simulate the performance of proposed architectures for HWO.}
The HPIC {is} publicly available to the community on the NASA Exoplanet Archive at { \url{https://exoplanetarchive.ipac.caltech.edu/docs/MissionStellar.html} \citep{HPIC_dataset}}.

\section*{Acknowledgements}
We thank Joshua Pepper and Keivan Stassun for providing information on the construction of the TIC and how its stellar properties could be updated in the HPIC. 
{We acknowledge Travis Berger for his advice on selecting the best sources of stellar astrophysical properties.}
NWT is supported by an appointment with the NASA Postdoctoral Program at the NASA Goddard Space Flight Center, administered by Oak Ridge Associated Universities under contract with NASA.
Part of this research was carried out in part at the Jet Propulsion Laboratory, California Institute of Technology, under a contract with the National Aeronautics and Space
Administration (80NM0018D0004).
This research has made us of the VizieR catalogue access tool, xMatch cross-match service, and the SIMBAD database, operated at CDS, Strasbourg, France.
This research has made use of the NASA Exoplanet Archive, which is operated by the California Institute of Technology, under contract with the National Aeronautics and Space Administration under the Exoplanet Exploration Program.
This work has made use of data from the European Space Agency (ESA) mission
{\it Gaia} (\url{https://www.cosmos.esa.int/gaia}), processed by the {\it Gaia}
Data Processing and Analysis Consortium (DPAC,
\url{https://www.cosmos.esa.int/web/gaia/dpac/consortium}). Funding for the DPAC
has been provided by national institutions, in particular the institutions
participating in the {\it Gaia} Multilateral Agreement.

\software{astropy \citep{astropy:2013, astropy:2018, astropy:2022},
astroquery \citep{Ginsburg2019}}

\appendix

\section{List of columns in the HPIC}
\label{HPIC_columns}
\startlongtable
\begin{deluxetable}{l|l}
\tablecaption{Description of columns in the HPIC}
\tablehead{\colhead{Name} & \colhead{Description}}
\startdata
        {star\_name} & unique object name in the HPIC \\
        {ra} & right ascension at epoch 2000 (ICRS) \\
        {dec} & declination at epoch 2000 (ICRS) \\
        ra\_J2016 & right ascension at epoch 2016 (ICRS) \\
        {dec\_J2016} & declination at epoch 2016 (ICRS) \\
        {tic\_id} & TESS Input Catalog ID \\
        {gaia\_dr2\_id} & Gaia DR2 ID \\
        {gaia\_dr3\_id} & Gaia DR3 ID \\ 
        {hip\_name} & Hipparcos ID \\ 
        {tm\_name} & 2MASS ID, add “2MASS J” \\
        {tyc\_name} & Tycho ID \\ 
        {wds\_designation} & Washington Double Star catalog ID \\
        {simbad\_name} & Name in the CDS Simbad database \\
        source\_list\_fl & Flag for the source of the object \\ 
        {sy\_pmra} & proper motion in RA (mas/yr) \\ 
        {sy\_pmraerr} & uncertainty in RA proper motion \\ 
        {sy\_pmdec} & proper motion in DEC (mas/yr) \\ 
        {sy\_pmdec} & uncertainty in DEC proper motion \\ 
        {sy\_pmsrc} & source of proper motion \\ 
        {sy\_pm\_reflink} & {proper motion bibcode} \\
        {sy\_ujmag} & Johnson $U$ mag \\ 
        {sy\_ujmagerr} & $U$ mag uncertainty \\ 
        {sy\_ujmagsrc}& $U$ mag source \\
        {sy\_ujmag\_reflink}   & {U mag bibcode} \\
        {sy\_bmag} & Johnson $B$ mag \\ 
        {sy\_bmagerr} & $B$ mag uncertainty \\ 
        {sy\_bmagsrc} & $B$ mag source \\ 
        {sy\_bmag\_reflink}   & {B mag bibcode} \\
        {sy\_vmag} & Johnson $V$ mag \\ 
        {sy\_vmagerr} & $V$ mag uncertainty \\ 
        {sy\_vmagsrc} & $V$ mag source \\ 
        {sy\_vmag\_reflink}   & $V$ mag bibcode \\
        {sy\_rcmag} & Johnson-Cousins $R$ mag \\ 
        {sy\_rcmagerr} & $R$ mag uncertainty \\ 
        {sy\_rcmagsrc} & $R$ mag source \\
        {sy\_rcmag\_reflink} & {$R$ mag bibcode}\\
        {sy\_icmag} & Johnson-Cousins $I$ mag \\ 
        {sy\_icmagerr} & $I$ mag uncertainty \\ 
        {sy\_icmagsrc} & $I$ mag source \\ 
        {sy\_icmag\_reflink} & {$I$ mag bibcode} \\
        {sy\_jmag} & 2MASS $J$ mag \\ 
        {sy\_jmagerr} & $J$ mag uncertainty \\ 
        {sy\_jmagsrc} & $J$ mag source \\ 
        {sy\_jmag\_reflink} & {$J$ mag bibcode} \\
        {sy\_hmag}& 2MASS $H$ mag \\ 
        {sy\_hmagerr} & $H$ mag uncertainty \\ 
        {sy\_hmagsrc} & $H$ mag source \\ 
        {sy\_hmag\_reflink} & {$H$ mag bibcode} \\
       {sy\_kmag} & 2MASS $K_s$ mag \\ 
        {sy\_kmagerr} & $K_s$ mag uncertainty \\ 
        {sy\_kmagsrc} & $K_s$ mag source \\ 
        {sy\_kmag\_reflink} & {$K_s$ mag bibcode}\\
        {sy\_tmag} & TESS mag \\ 
        {sy\_tmagerr} & TESS mag uncertainty \\ 
        {sy\_gaiamag} & Gaia magnitude \\ 
        {sy\_gaiamagerr} & $G$ mag uncertainty \\ 
        {sy\_gaiamagsrc} & $G$ mag source (always same source as $G_{BP}$, $G_{RP}$ when available)\\ 
        {sy\_gaiamag\_reflink} & {$G$ mag bibcode} \\
        {sy\_bpmag} & Gaia $G_{BP}$ mag \\ 
        {sy\_bpmagerr} & $G_{BP}$ mag uncertainty \\ 
        {sy\_rpmag} & Gaia $G_{RP}$ mag \\ 
        {sy\_rpmagerr} & $G_{RP}$ mag uncertainty \\ 
        {sy\_plx} & parallax (mas) \\ 
        {sy\_plxerr} & parallax uncertainty \\ 
        {sy\_plxsrc} & parallax source \\
        {sy\_plx\_reflink} & {parallax bibcode} \\
        {sy\_dist} & Distance (pc) \\
        {sy\_disterr} & distance uncertainty \\ 
        {sy\_distsrc} & distance source \\
        {sy\_dist\_reflink} & {distance bibcode} \\
        ambiguous\_dist\_fl & flag for distance measurements that deviate from the value given by Simbad \\
        {st\_spectype} & Spectral type from simbad \\ 
        {st\_spectype\_reflink} & bibcode of spectral type \\ 
        dwarf\_fl & flag for if object is a dwarf star \\ 
        {st\_teff} & Stellar effective temperature (K) \\ 
        {st\_tefferr} & $\mathrm{T_{eff}}$ uncertainty \\ 
        {st\_teffsrc} & $\mathrm{T_{eff}}$ source \\
        {st\_teff\_reflink} & {$\mathrm{T_{eff}}$ bibcode} \\
        {st\_logg} & stellar log surface gravity (from bulk properties, log cgs units) \\ 
        {st\_loggerr} & bulk log(g) uncertainty \\
        {st\_loggsrc} & bulk log(g) source \\
        {st\_logg\_reflink} & { bulk log(g) bibcode} \\
        {st\_loggspec} & stellar log surface gravity (from spectroscopy, log cgs units) \\ 
        {st\_loggspecerr} & spec log(g) uncertainty \\
        {st\_loggspecsrc} & spec log(g) source \\
        {st\_loggspec\_reflink} & {spec log(g) bibcode} \\
        {st\_met} & stellar metallicity in [Fe/H] \\ 
        {st\_meterr} & [Fe/H] uncertainty \\
        {st\_metsrc} & [Fe/H] source \\
        {st\_met\_reflink} & {[Fe/H] bibcode} \\
        {st\_lum} & $\log_{10}$ stellar luminosity ($L_{\odot}$) \\ 
        {st\_lumerr} & $\log_{10}$ luminosity uncertainty \\ 
        {st\_lumsrc} & luminosity source \\
        {st\_lum\_reflink} & {log(L) bibcode} \\
        {st\_rad} & stellar radius ($R_{\odot}$) \\ 
        {st\_raderr} & radius uncertainty \\ 
        {st\_radsrc} & radius source \\
        {st\_rad\_reflink} & {radius bibcode} \\
        {st\_mass} & stellar mass ($M_{\odot}$) \\ 
        {st\_masserr} & mass uncertainty \\ 
        {st\_masssrc} & mass source \\ 
        {st\_mass\_reflink} & {mass bibcode} \\
        uncertain\_M\_flag & flag for if mass measurement may be unreliable \\ 
        {st\_age} & stellar age (Gyr) \\ 
        {st\_ageerr} & age uncertainty \\ 
        {st\_agesrc} & age source \\ 
        {st\_age\_reflink} & {age bibcode} \\
        contaminant\_fl & flag if there are nearby contaminants \\
        brightest\_sep & angular separation from brightest Gaia contaminant (") \\
        brightest\_Gmag & $G$ magnitude difference with brightest Gaia contaminant \\ 
        brightest\_id & Gaia id of brightest contaminant \\
        nearest\_sep & angular separation from nearest Gaia contaminant (") \\
        nearest\_Gmag & $G$ magnitude difference with nearest Gaia contaminant \\
        nearest\_id & Gaia id of nearest contaminant \\
        known\_binary\_fl & flag if object is a known binary \\ 
        gaia\_binary\_fl & flag if object is tagged as binary by gaia dr3 \\
        {wds\_comp} & {WDS system component} \\
        {wds\_sep} & WDS angular separation (") \\ 
        {wds\_delta\_mag} & WDS magnitude difference \\
        GCNS\_companion & Gaia DR3 id for object’s companion in Gaia Catalog of Nearby Stars \\ 
        GCNS\_sep & GCNS separation (") \\
        GCNS\_mag\_diff & GCNS $G$ magnitude difference \\ 
        {sy\_planets\_flag} & flag for if there are known planets \\
        {hostname} & name of the exoplanet host star in the NASA Exoplanet Archive \\ \hline
\enddata
\end{deluxetable}

\section{List of HPIC flags}
\label{HPIC_flags}
\startlongtable
\begin{deluxetable}{lll}
\tablewidth{8in}
\tablecaption{List of source and quality flags in the HPIC}
\tablehead{\colhead{Name} & \colhead{Flag} & \colhead{Description}}
\startdata
source\_list\_fl & ~ & Flag for the source catalog of an object \\ 
        ~ & tic\_near & Nearby (dist $<$ 50pc) objects in the TIC \\ 
        ~ & tic\_bright & Bright ($T<8$) objects in the TIC \\ 
        ~ & gaia\_dr3 & objects from Gaia DR3 \\ 
        {sy\_pmsrc} & ~ & source flag for the proper motion \\ 
        ~ & GaiaDR3 & from Gaia DR3 \\ 
        ~ & TIC  \{source\} & from TIC, with source flag \{source\} \\ 
        ~ & Simbad  \{source\} & from Simbad with bibcode \{source\} \\ 
        {sy\_\{band\}src} & ~ & source of the magnitude in photometric band \{band\} \\ 
        ~ & TIC \{source\} & from TIC, with source flag \{source\} \\ 
        ~ & gaia\_dr3 & from Gaia DR3 \\ 
        ~ & Simbad  \{source\} & from Simbad with bibcode \{source\} \\ 
        ~ & gaia10pc & From 10 parsec sample of \citet{Reyle2021} \\ 
        ~ & GCNS & from Gaia Catalog of Nearby Stars \\ 
        {sy\_plxsrc} & ~ & source of the parallax \\ 
        ~ & GaiaDR3 & from Gaia DR3 \\ 
        ~ & TIC \{source\} & from TIC, with source flag \{source\} \\ 
        ~ & Simbad  \{source\} & from Simbad with bibcode \{source\} \\ 
        {sy\_distsrc} & ~ & source of the Distance \\ 
        ~ & bj2021 & Gaia DR3 distance estimate from \citet{bailerjones2021} \\ 
        ~ & TIC \{source\} & from TIC, with source flag \{source\} \\ 
        ~ & Simbad  \{source\} & from Simbad with bibcode \{source\} \\ 
        ~ & gaia\_plx & calculated using gaia parallax \\ 
        ~ & TIC plx \{source\} & calculated from parallax given in TIC with source flag \{source\} \\ 
        ~ & Simbad plx \{source\} & calculated from parallax given by Simbad with bibcode \{source\} \\ 
        ~ & from\_companion & from distance of known binary companion \\ 
        ambiguous\_dist\_fl & ~ & boolean flag for distance measurements that deviate from the value given by Simbad \\  
        dwarf\_fl & ~ & Boolean flag for if object is a dwarf star \\ 
        {st\_teffsrc} & ~ & source of the effective temperature \\ 
        ~ & Pastel & from \citet{soubiran2022} \\ 
        ~ & TIC & from TIC  \\ 
        ~ & gspspec & from Gaia GSP-Spec pipeline \\ 
        ~ & gspphot & from Gaia GSP-Phot pipeline \\ 
        ~ & gcs\_reanalysis & from Geneva-Copenhagen Survey reanalysis of \citet{casagrande2011} \\ 
        ~ & Simbad  \{source\} & from Simbad with bibcode \{source\} \\ 
        ~ & McDonald2017 & from \citet{mcdonald2017} \\ 
        ~ & using\_TIC\_relation & calculated using the empirical relation in \citet{stassun2019} \\ 
        ~ & Pecaut\_and\_Mamajek\_2013 & calculated using the B-V color in the \citet{pecaut2013} empirical relation \\ 
        {st\_loggsrc} & ~ & source of log(g) value, same flags as st\_loggspecsrc \\ 
        ~ & Pastel & from \citet{soubiran2022} \\ 
        ~ & gspspec & from Gaia GSP-Spec pipeline \\ 
        ~ & gspphot & from Gaia GSP-Phot pipeline \\ 
        ~ & gcs\_reanalysis & from Geneva-Copenhagen Survey reanalysis of Casagrande+, 2011 \\ 
        ~ & Simbad  \{source\} & from Simbad with bibcode \{source\} \\ 
        ~ & McDonald2017 & from \citet{mcdonald2017} \\ 
        ~ & Stock2018 & from \citet{stock2018} catalog of giant star properties \\ 
        ~ & calculated\_from\_M\_and\_R & calculated using known stellar mass and radius \\ 
        {st\_metsrc} & ~ & source of metallicity \\ 
        ~ & Pastel & from \citet{soubiran2022}\\ 
        ~ & gspspec [M/H] & from Gaia GSP-Spec pipeline \\ 
        ~ & gspphot [M/H] & from Gaia GSP-Phot pipeline \\ 
        ~ & gcs\_reanalysis & from Geneva-Copenhagen Survey reanalysis of \citet{casagrande2011} \\ 
        ~ & TIC & from TIC \\ 
        ~ & Simbad  \{source\} & from Simbad with bibcode \{source\} \\ 
        {st\_lumsrc} & ~ & source of luminosity \\ 
        ~ & update\_TIC & calculated using the value in the TIC and with updated distance \\ 
        ~ & update\_TIC\_gaia2 & calculated updating TIC value derived from Gaia DR2 bolometric correction \\ 
        ~ & calc\_from\_gaia & calculated using Gaia bolometric correction \\ 
        ~ & gaia\_dr3\_flame & calculated via Gaia FLAME pipeline \\ 
        ~ & gcs\_Fbol & calculated using the bolometric flux in \citet{casagrande2011} \\ 
        ~ & Pecaut\_and\_Mamajek\_2013 & using $B-V$ relation in \citet{pecaut2013} \\ 
        ~ & update McDonald2017   & updating luminosity given in \citet{mcdonald2017} using current distance \\ 
        ~ & McDonald2017 & from \citet{mcdonald2017} \\ 
        ~ & Stock2018 & from \citet{stock2018} catalog of giant star properties \\ 
        {st\_radsrc} & ~ & source of radius \\ 
        ~ & update\_TIC & calculated using the value in the TIC and with updated distance \\ 
        ~ & update\_TIC\_gaia2 & calculated updating TIC value derived from Gaia DR2 bolometric correction \\ 
        ~ & calc\_from\_gaia & calculated using Gaia bolometric correction \\ 
        ~ & gaia\_dr3\_flame & calculated via Gaia FLAME pipeline \\ 
        ~ & McDonald2017 & from \citet{mcdonald2017} \\ 
        ~ & Stock2018 & from \citet{stock2018} catalog of giant star properties \\ 
        ~ & calc\_from\_L\_and\_Teff & calculated using the Stefan-Boltzmann relation \\ 
        {st\_masssrc} & ~ & source of mass \\ 
        ~ & Mann2019 & using \citet{mann2019} empirical relation \\ 
        ~ & gaia\_dr3\_flame & calculated via Gaia FLAME pipeline \\ 
        ~ & TIC & from TIC \\ 
        ~ & TIC\_teff\_mass\_relation & using empirical relation in \citet{stassun2019} \\ 
        ~ & gcs\_reanalysis & from Geneva-Copenhagen Survey reanalysis of \citet{casagrande2011} \\ 
        ~ & Kordopatis2023 & from isochrones of \citet{kordopatis2023} \\ 
        ~ & Stock2018 & from \citet{stock2018} catalog of giant star properties \\ 
        ~ & JimenezEsteban2023 & from \citet{Jimenez2023} catalog of white dwarfs \\ 
        uncertain\_M\_flag & ~ & Boolean flag for if mass measurement may be unreliable \\ 
        {st\_agesrc} & ~ & source of age \\ 
        ~ & gaia\_dr3\_flame & calculated via Gaia FLAME pipeline \\ 
        ~ & gcs\_reanalysis & from Geneva-Copenhagen Survey reanalysis of \citet{casagrande2011} \\ 
        ~ & Kordopatis2023 & from isochrones of \citet{kordopatis2023} \\ 
        ~ & Stock2018 & from \citet{stock2018} catalog of giant star properties \\ 
        contaminant\_fl & ~ & Boolean flag if there are nearby contaminants \\ 
        known\_binary\_fl & ~ & Boolean flag if object is a known binary \\ 
        gaia\_binary\_fl & ~ & Boolean flag if object is tagged as binary by Gaia DR3, \\
        ~ & ~ & includes poorly characterized spectroscopic and astrometric binaries \\ 
        {sy\_planets\_flag} & ~ & Boolean flag for if there are known planets \\ \hline
\enddata
\end{deluxetable}

\bibliographystyle{aasjournal}
\bibliography{sources}
\end{document}